\documentclass{PoS}
\usepackage{algorithmicx,algorithm,algpseudocode}

\title{Exploring free-form smearing for bottomonium and B meson spectroscopy}

\ShortTitle{Exploring free-form smearing for bottomonium and B meson spectroscopy}

\author{Mark Wurtz,$^a$ Randy Lewis,$^a$ and \speaker{R. M. Woloshyn},$^b$ \\
        \llap{$^a$} Department of Physics and Astronomy, York University, Toronto, ON, M3J 1P3, Canada \\
        \llap{$^b$} TRIUMF, 4004 Wesbrook Mall, Vancouver, BC, V6T 2A3, Canada \\
        E-mail: \email{mwurtz@yorku.ca}, \email{randy.lewis@yorku.ca}, \email{rwww@triumf.ca}}

\abstract{Free-form smearing was designed as a way to implement source operators of any desired shape. A variation of the method is introduced that reduces the computational cost by reducing the number of link multiplications to its absolute minimum. Practical utility is
demonstrated through calculations of bottomonium and B meson masses.}

\FullConference{The 33rd International Symposium on Lattice Field Theory\\
		 14 -18 July  2015\\
		 Kobe International Conference Center, Kobe, Japan}

\begin{document}

\section{Motivation}

To study a particular hadronic state, it would be ideal to have an
operator that couples only to that state. An operator with the correct
angular momentum quantum numbers is necessary but not sufficient. The
spatial structure of the operator (i.e.\ its ``shape'') helps to distinguish states
with different principal quantum numbers.

Free-form smearing was designed as a way to implement source operators of any desired shape \cite{vonHippel:2013yfa} and
a variation of the method has been introduced that reduces the computational cost by reducing the number of link multiplications to its absolute minimum
\cite{Wurtz:2015mqa}.

An explicit implementation of the new method is provided here, and
the practical utility of the algorithm is demonstrated through calculations of bottomonium and $B_c$, $B_s$ and $B$ meson masses.

\section{Free-form smearing}\label{ffs}

Though free-form smearing can be applied to a quark propagator in any context,
consider for definiteness a meson operator defined at a lattice site $x$.
Let $\chi(x)$ represent the antiquark field and let $\tilde{\tilde\psi}(x)$ represent the free-form smeared quark field.
The meson operator is then
\begin{equation}\label{mesonop}
\chi(x)\tilde{\tilde\psi}(x) = \chi(x)\sum_y\Omega(x-y)\frac{\tilde{\psi}_x(y)}{\left<\left|\left|\tilde{\psi}_x(y)\right|\right|\right>}
\end{equation}
where $\Omega(x-y)$ is chosen by the user to define the desired shape and quantum numbers of the operator.
The denominator is the ensemble average of a color+Dirac trace
\begin{equation}
\left<\left|\left|\tilde{\psi}_x(y)\right|\right|\right> = \sqrt{{\rm Tr}\left(\tilde{\psi}_x^\dagger(y)\tilde{\psi}_x(y)\right)}
\end{equation}
and its purpose is to divide out spatial variations in the quark field $\tilde{\psi}_x(y)$
so the interpretation of $\Omega(x-y)$ is as transparent as possible.

The original implementation of free-form smearing defined $\tilde\psi$ through
standard Gaussian smearing,
but a more efficient definition is the minimal-path definition, where no
paths are duplicated in the calculation:
\begin{equation}
\tilde{\psi}_x(y) = \sum_{\rm shortest~paths}U(x\to y)\psi(y)
\end{equation}
The link variables in $U(x\to y)$ can be thin or thick; our
examples use stout links \cite{Morningstar:2003gk}.

\section{The minimal-path algorithm to compute \boldmath{$\tilde\psi_x(y)$}}

The algorithm to compute the minimal paths of links is presented as pseudocode in Alg.~\ref{alg}. Free-form smearing can then be implemented as described in Sec.~\ref{ffs}.
Whereas the Gaussian version of free-form smearing requires $O(L^4)$ link multiplications,
the minimal-path method uses only $3L^3$ link multiplications, which is exactly the number of spatial links on a given time slice and is therefore the minimal number of link multiplications.

\begin{algorithm}
\caption{An implementation of the minimal-path algorithm for free-form smearing.}\label{alg}
\begin{algorithmic}
\color{blue}
\State sitevisited = false \Comment{\color{black}Initialize all sites and links as having not been visited.}
\color{blue}
\State linkvisited = false
\color{blue}
\State $\tilde{\psi}=0$ \Comment{\color{black}Initialize the field to unity at the origin and zero everywhere else.}
\color{blue}
\State $\tilde{\psi}(x)=1$
\color{blue}
\State sitevisited$(x)$ = true \Comment{\color{black}Mark the origin, $x$, as having been visited.}
\color{blue}
\State $y_{\rm frontier}(1)=x$ \Comment{\color{black}Mark the origin, $x$, as on the frontier.}
\color{blue}
\State $n=1$ \Comment{\color{black}Initially there is only one point on the frontier.}
\color{blue}
\While{$n>0$} \Comment{\color{black}Loop until all sites are visited, \textit{i.e.} the frontier contains no points $(n=0)$.}
\color{blue}
  \State $n_{\rm new}=0$ \Comment{\color{black}Initialize the number of points on the new frontier to zero.}
\color{blue}
  \For {$i = 1,n$} \Comment{\color{black}Visit all of the points on the frontier.}
\color{blue}
  \State $y = y_{\rm frontier}(i)$
\color{blue}
    \For {$\mu=1,3$}
\color{blue}
      \State \Comment{\color{black}FORWARD DIRECTION:}
\color{blue}
      \If {linkvisited$(\mu,y)$ = false} \Comment{\color{black}If the link $U_\mu(y)$ has not been visited.}
\color{blue}
        \State $\tilde{\psi}(y+\mu) = \tilde{\psi}(y+\mu) + U^\dag_\mu(y) \tilde{\psi}(y)$ \Comment{\color{black}Use $U_\mu(y)$ to update the new frontier.}
\color{blue}
        \State linkvisited$(\mu,y)$ = true \Comment{\color{black}Mark this link as visited.}
\color{blue}
        \If {sitevisited$(y+\mu)$ = false} \Comment{\color{black}If this site has not been visited before.}
\color{blue}
          \State sitevisited$(y+\mu)$ = true \Comment{\color{black}Mark this site as visited.}
\color{blue}
          \State $n_{\rm new} = n_{\rm new} + 1$ \Comment{\color{black}The new frontier is larger.}
\color{blue}
          \State $y_{\rm new frontier}(n_{\rm new}) = y+\mu$ \Comment{\color{black}Add this site to the new frontier.}
\color{blue}
        \EndIf
\color{blue}
      \EndIf
\color{blue}
      \State \Comment{\color{black}BACKWARD DIRECTION:}
\color{blue}
      \If {linkvisited$(\mu,y)$ = false} \Comment{\color{black}If the link $U_\mu(y-\mu)$ has not been visited.}
\color{blue}
        \State $\tilde{\psi}(y-\mu) = \tilde{\psi}(y-\mu) + U_\mu(y-\mu) \tilde{\psi}(y)$ \Comment{\color{black}Use $U_\mu(y-\mu)$ to update new frontier.}
\color{blue}
        \State linkvisited$(\mu,y)$ = true \Comment{\color{black}Mark this link as visited.}
\color{blue}
        \If {sitevisited$(y-\mu)$ = false} \Comment{\color{black}If this site has not been visited before.}
\color{blue}
          \State sitevisited$(y-\mu)$ = true \Comment{\color{black}Mark this site as visited.}
\color{blue}
          \State $n_{\rm new}=n_{\rm new} + 1$ \Comment{\color{black}The new frontier is larger.}
\color{blue}
          \State $y_{\rm new frontier}(n_{\rm new})=y-\mu$ \Comment{\color{black}Add this site to the new frontier.}
\color{blue}
        \EndIf
\color{blue}
      \EndIf
\color{blue}
    \EndFor
\color{blue}
  \EndFor
\color{blue}
  \State $n=n_{\rm new}$ \Comment{\color{black}Sites in the new frontier become sites in the old frontier.}
\color{blue}
  \State $y_{\rm frontier}=y_{\rm new frontier}$
\color{blue}
\EndWhile
\end{algorithmic}
\end{algorithm}

\section{Sample smearing shapes}

To smear a bottom quark for use in bottomonium or a bottom-flavored meson,
we choose $\Omega(x-y)$ to have a Coulomb wave function shape (familiar from the quantum
mechanics of hydrogen) as well as the requisite $J^P$
quantum numbers.  All of the $\Omega(x-y)$ functions from S wave to G wave are
listed in \cite{Wurtz:2015mqa} so only the first few are displayed here, in Table~\ref{shapetable}.
\begin{table}
\caption{The $\Omega(x-y)$ functions of Eq.~(\protect\ref{mesonop}) that represent Coulomb wave function shapes for S-wave and P-wave mesons can be expressed as $\Omega(x-y)=e^{-r/a_0}f(x-y)$.  A few of the $f(x-y)$ are tabulated here.}\label{shapetable}
\begin{center}
\begin{tabular}{rlll}
\hline\hline
$^{2S+1}L_J$ & 0th radial & 1st radial & 2nd radial \\
\hline
$^1S_0$ & 1 & $(r-b)$ & $(r-c)(r-b)$ \\
$^3S_1$ & $\sigma_i$ & $(r-b)\sigma_i$ & $(r-c)(r-b)\sigma_i$ \\
$^1P_1$ & $\sin\left(\frac{2\pi}{L}(x_i-y_i)\right)$ & $(r-b)\sin\left(\frac{2\pi}{L}(x_i-y_i)\right)$ \\
$^3P_0$ & $\sum_{i=1}^3\sigma_i\sin\left(\frac{2\pi}{L}(x_i-y_i)\right)$ & $(r-b)\sum_{i=1}^3\sigma_i\sin\left(\frac{2\pi}{L}(x_i-y_i)\right)$ \\
\hline\hline
\end{tabular}
\end{center}
\end{table}

The radius and nodal parameters $(a_0,b,c)$ are tuned for each hadron to optimize the signal, and in all of the studies
reported here they satisfy
$0.5\leq a_0\leq7.0$ and $2.1<b\leq6.2$ and $c=6.0$ in lattice units.
These parameters are determined quite precisely for each meson, and the ranges listed
here are only to provide a notion of typical numerical values.
The argument $r$ of $\Omega(r)$ is defined by
\begin{equation}
r = \sqrt{(x_1-y_1)_{\rm min}^2+(x_2-y_2)_{\rm min}^2+(x_3-y_3)_{\rm min}^2}.
\end{equation}

\section{Simulation details}

The spectrum calculations presented here use an ensemble of 198 configurations produced by the PACS-CS collaboration \cite{Aoki:2008sm}.
Bottom quark propagators are computed with $O(v^4)$ tadpole-improved lattice NRQCD.
All other quark propagators are
clover-improved Wilson fermions computed using the sap\_gcr solver from DD-HMC \cite{Luscher:2005rx}.
Parameters for the charm quark, following the Fermilab interpretation, are from \cite{Lang:2014yfa}.

Three methods were combined to increase statistical precision:
(1)
a random U(1) wall source was used, with support from $4^3$ evenly spaced lattice sites,
(2)
correlators were averaged over sources on different time slices (every time step for bottomonium or every second time step for bottom mesons),
(3)
NRQCD propagators were calculated forward and backward in time.

\section{Results}

\begin{figure}
\begin{center}
\includegraphics[width=12cm,angle=0,trim=0 0 0 0,clip=true]{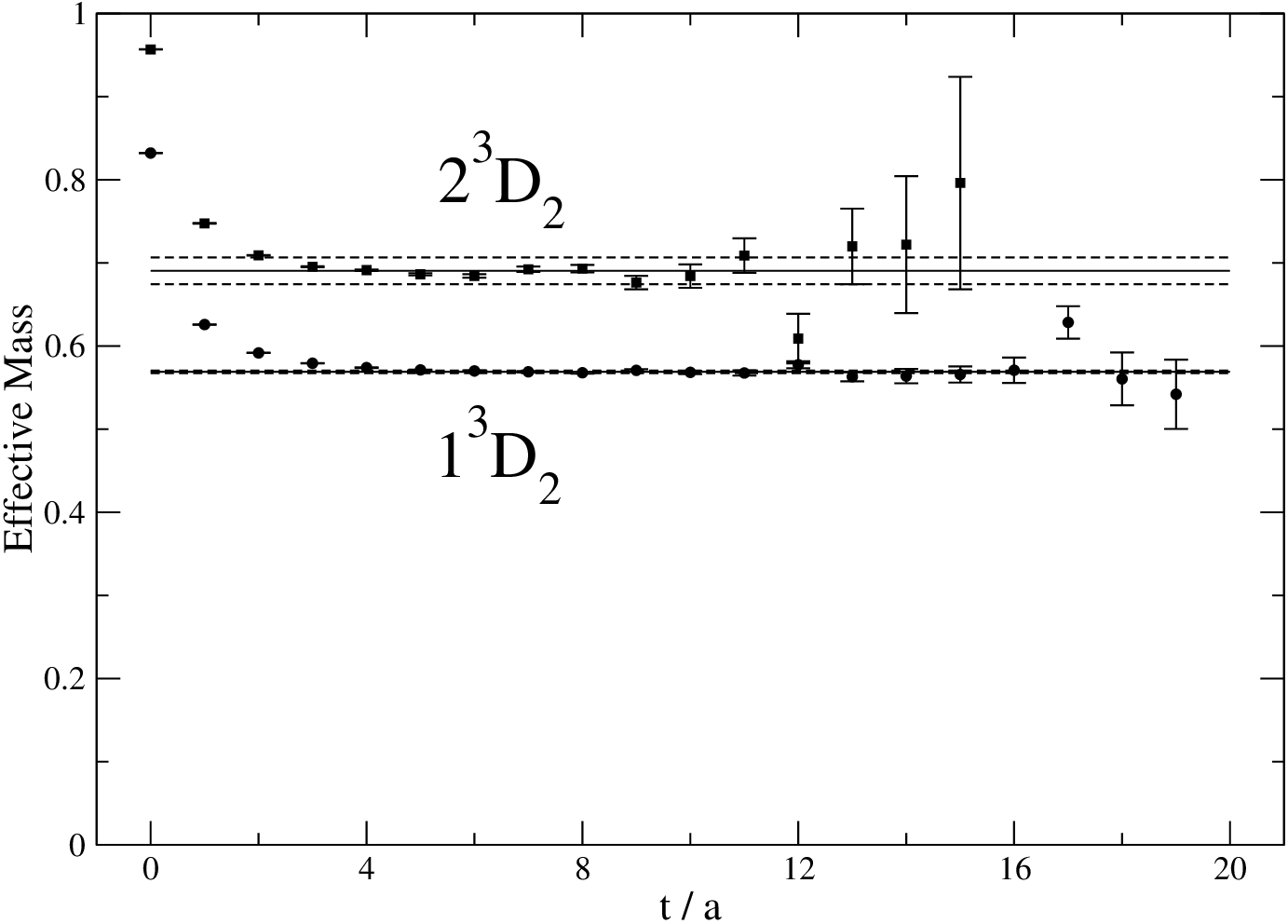}
\end{center}
\caption{Effective mass plots demonstrating that the $n=2$ operator is orthogonal
to the $n=1$ operator for $^3D_2$ bottomonium.}\label{fig:orthog}
\end{figure}

To verify orthogonality based on operator shape, Fig.~\ref{fig:orthog} shows
effective mass plots for the $^3D_2$ bottomonium operators for the zero'th and first
radial excitations.
Notice that the $n$=2 result shows no contamination from the lighter $n$=1 meson.
This success is particularly satisfying because no signal for $2^3D_2$ bottomonium has been reported in the lattice QCD literature before this work.

\begin{figure}
\begin{center}
\includegraphics[width=12cm,angle=0,trim=0 0 0 0,clip=true]{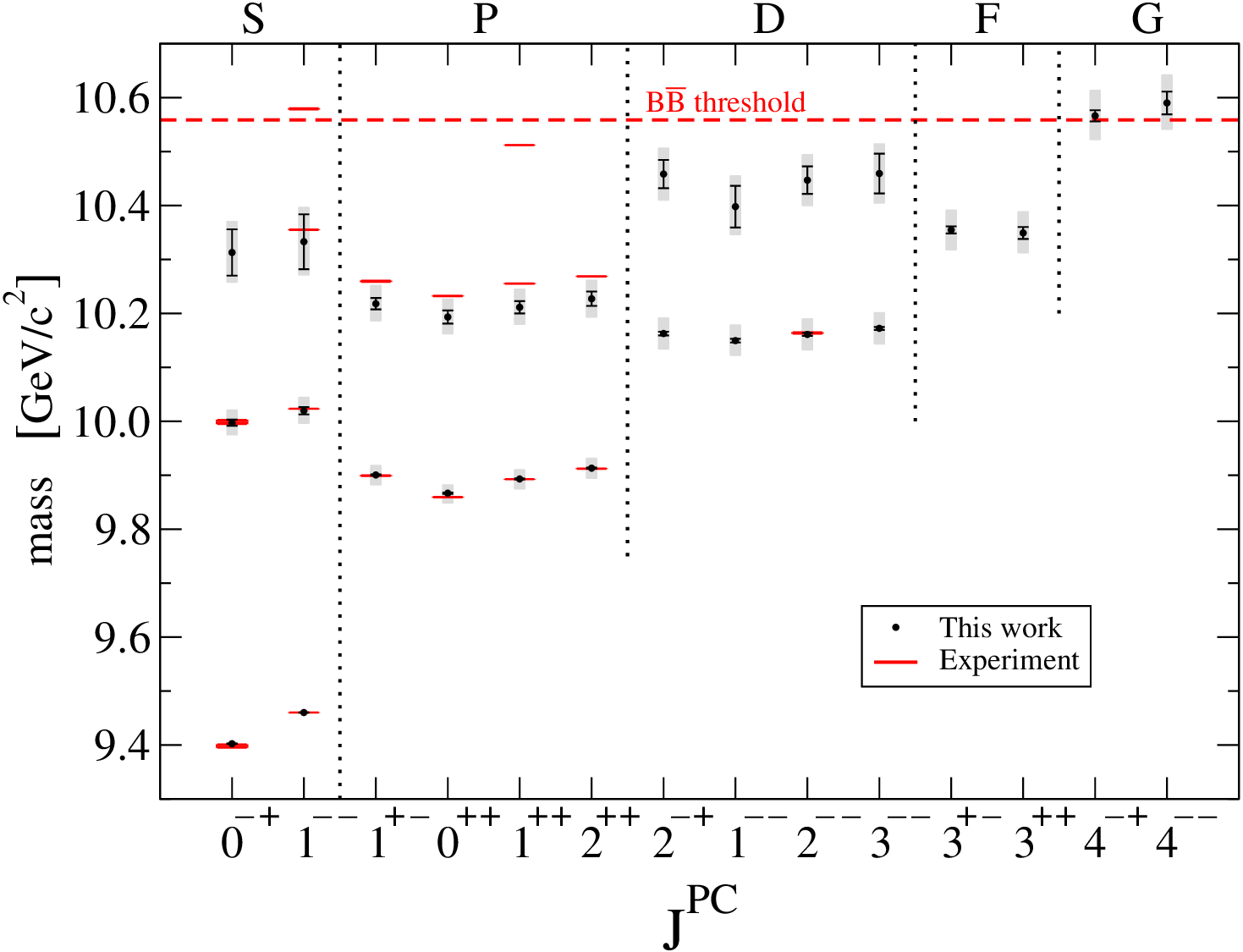}
\end{center}
\caption{The bottomonium spectrum obtained using minimal-path free-form smearing.}\label{fig:BB}
\end{figure}

\begin{figure}
\begin{center}
\includegraphics[width=12cm,angle=0,trim=0 0 0 0,clip=true]{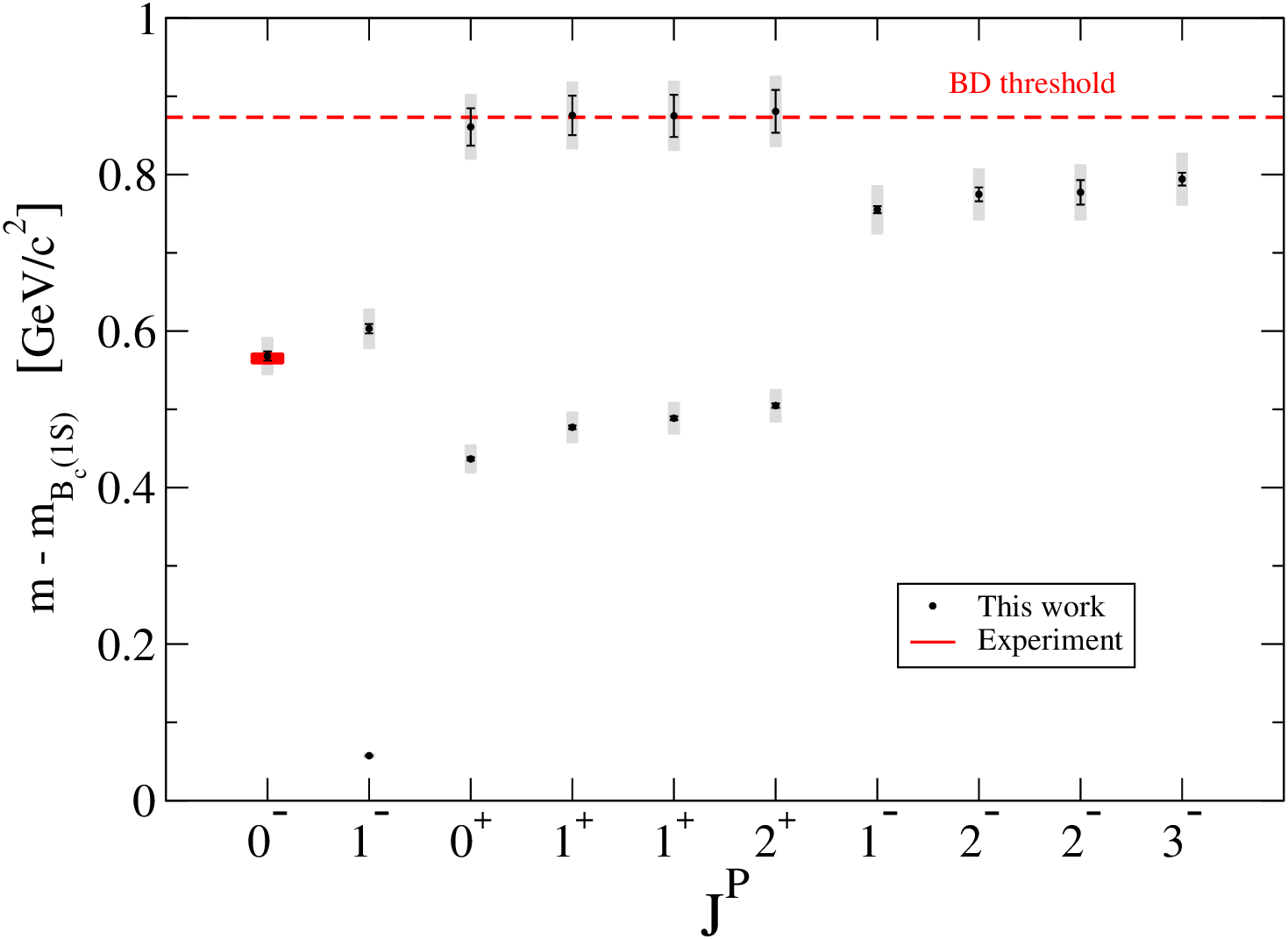}
\end{center}
\caption{The $B_c$ meson spectrum obtained using minimal-path free-form smearing.}\label{fig:Bc}
\end{figure}

\begin{figure}
\begin{center}
\includegraphics[width=12cm,angle=0,trim=0 0 0 0,clip=true]{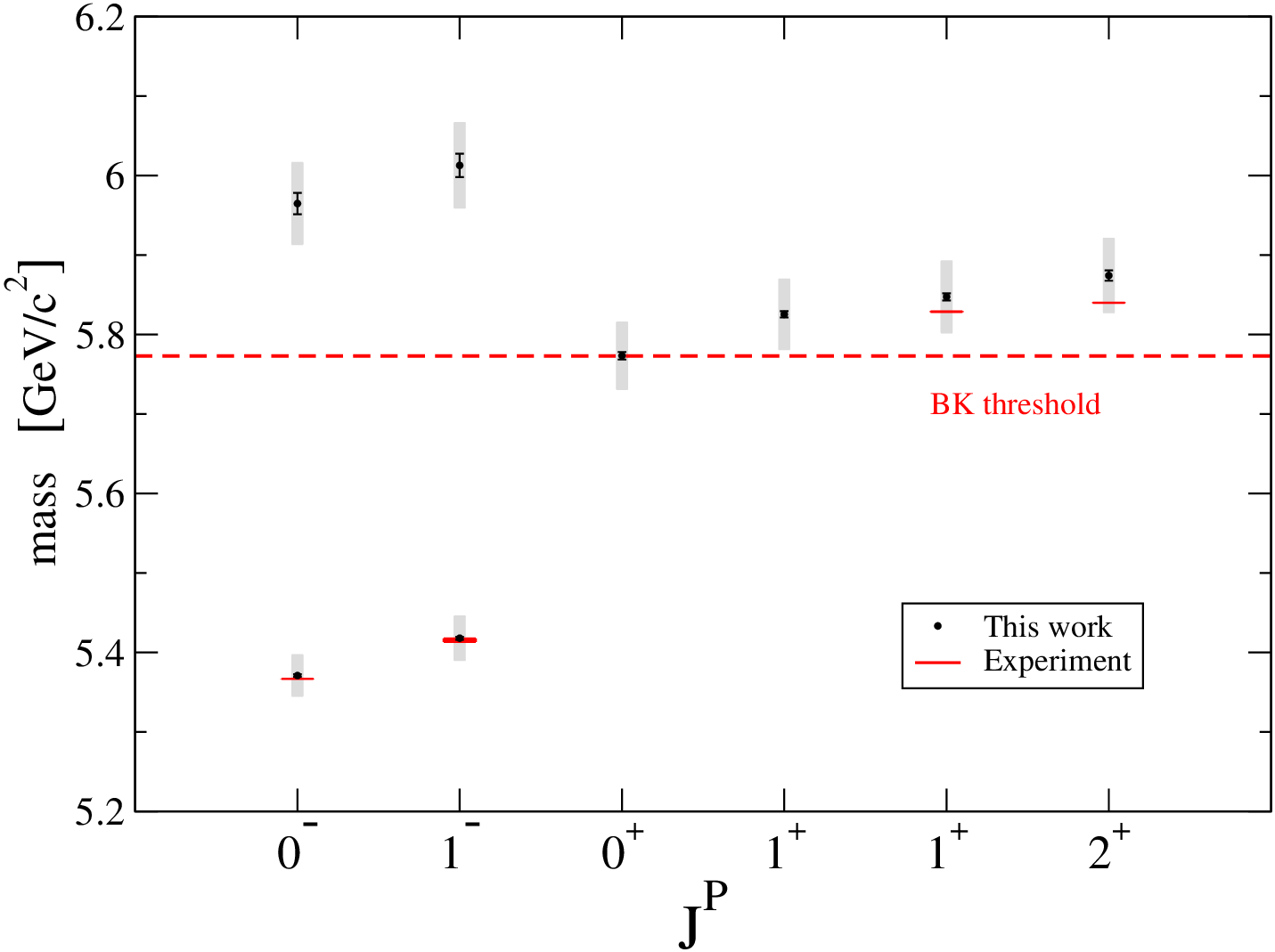}
\end{center}
\caption{The $B_s$ meson spectrum obtained using minimal-path free-form smearing.}\label{fig:Bs}
\end{figure}

\begin{figure}
\begin{center}
\includegraphics[width=12cm,angle=0,trim=0 0 0 0,clip=true]{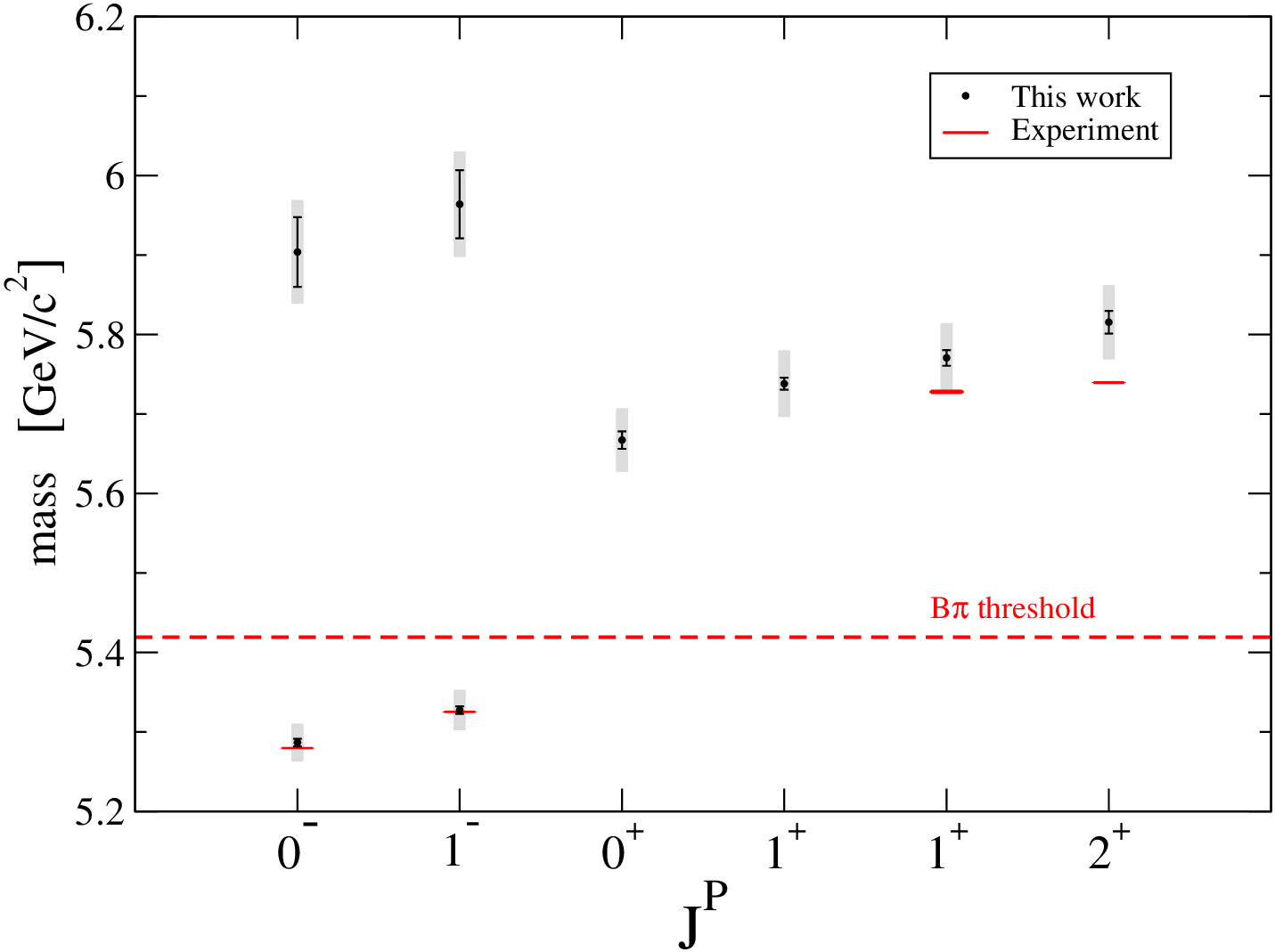}
\end{center}
\caption{The $B$ meson spectrum obtained using minimal-path free-form smearing.}\label{fig:B}
\end{figure}

To determine the complete spectrum, several operators were used for each $J^{PC}$ channel and a simultaneous fit was performed.
Summary plots are shown in Figs.~\ref{fig:BB}, \ref{fig:Bc}, \ref{fig:Bs}, and \ref{fig:B} and include experimental data \cite{Agashe:2014kda,Aaltonen:2013atp,Aad:2014laa,Aaij:2014hla,Aaij:2015qla} where available.
For further details, please consult \cite{Wurtz:2015mqa}.

\section*{Acknowledgments}
This work was supported in part by the Natural Sciences and Engineering Research
Council (NSERC) of Canada.


\begin{thebibliography}{99}
\bibitem{vonHippel:2013yfa} 
  G.~M.~von Hippel, B.~J\"ager, T.~D.~Rae and H.~Wittig,
  JHEP {\bf 1309}, 014 (2013).
\bibitem{Wurtz:2015mqa} 
  M.~Wurtz, R.~Lewis and R.~M.~Woloshyn,
  Phys.\ Rev.\ D {\bf 92}, 054504 (2015).
\bibitem{Morningstar:2003gk} 
  C.~Morningstar and M.~J.~Peardon,
  Phys.\ Rev.\ D {\bf 69}, 054501 (2004).
\bibitem{Aoki:2008sm} 
  S.~Aoki {\it et al.} [PACS-CS Collaboration],
  Phys.\ Rev.\ D {\bf 79}, 034503 (2009).
\bibitem{Luscher:2005rx} 
  M.~L\"uscher,
  Comput.\ Phys.\ Commun.\  {\bf 165}, 199 (2005).
\bibitem{Lang:2014yfa} 
  C.~B.~Lang, L.~Leskovec, D.~Mohler, S.~Prelovsek and R.~M.~Woloshyn,
  Phys.\ Rev.\ D {\bf 90}, 034510 (2014).
\bibitem{Agashe:2014kda} 
  K.~A.~Olive {\it et al.} [Particle Data Group Collaboration],
  Chin.\ Phys.\ C {\bf 38}, 090001 (2014).
\bibitem{Aaltonen:2013atp} 
  T.~A.~Aaltonen {\it et al.} [CDF Collaboration],
  Phys.\ Rev.\ D {\bf 90}, 012013 (2014).
\bibitem{Aad:2014laa} 
  G.~Aad {\it et al.} [ATLAS Collaboration],
  Phys.\ Rev.\ Lett.\  {\bf 113}, 212004 (2014).
\bibitem{Aaij:2014hla} 
  R.~Aaij {\it et al.} [LHCb Collaboration],
  JHEP {\bf 1410}, 88 (2014).
\bibitem{Aaij:2015qla} 
  R.~Aaij {\it et al.} [LHCb Collaboration],
  JHEP {\bf 1504}, 024 (2015).
\end{thebibliography}
\end{document}